\input{aipcheck}
\documentclass[
    ,final            
  ]
  {aipproc}
\layoutstyle{8x11single}

\newcommand{{\apj}}{{Astrophys. J}}

\def\la{{\le}}
\def\ga{{\ge}}
\newcommand{\rem}[1]{ }
\newcommand{\bea}{\begin{eqnarray}}
\newcommand{\eea}{\end{eqnarray}}
\newcommand{\beq}{\begin{equation}}
\newcommand{\eeq}{\end{equation}}

\def\apj{{\em Astrophys. J.} }
\def\apjl{{\em Astrophys. J. Lett.} }
\def\apjs{{ApJS} }

\begin{document}

\title{GRB physics with Fermi}

\classification{98.70.Rz, 52.35.Tc, 95.30.Qd, 98.70.Sa, 95.30.Gv}
\keywords      {gamma-ray bursts, shocks, magnetic fields, particle acceleration,  radiation}

\author{Mikhail V. Medvedev}{
  address={Department of Physics and Astronomy, University of Kansas, Lawrence, KS 66045, USA}
    ,altaddress={Russian Research Center "Kurchatov Institute", 123182 Moscow, Russia} 
}

\begin{abstract}
Radiation from GRBs in the prompt phase, flares and an afterglow is thought to be produced by accelerated electrons in magnetic fields. Such emission may be produced at collisionless shocks of baryonic outflows or at reconnection sites (at least for the prompt and flares) of the magnetically dominated (Poynting flux driven) outflows, where no shocks presumably form at all. An astonishing recent discovery is that during reconnection strong small-scale magnetic fields are produced via the Weibel instability, very much like they are produced at relativistic shocks. The relevant physics has been successfully and extensively studied with the PIC simulations in 2D and, to some extent, in 3D for the past few years. We discuss how these simulations predict the existence of MeV-range synchrotron/jitter emission in some GRBs, which can be observed with Fermi. Recent results on modeling of the spectral variability and spectral correlations of the GRB prompt emission in the Weibel-jitter paradigm applicable to both baryonic and magnetic-dominated outflows is reviewed with the emphasis on observational predictions. 
\end{abstract}

\maketitle

\section{Introduction}

Gamma-ray bursts --- the most luminous sources in the universe --- are conventionally explained as violent explosive or merger phenomena that drive a non-uniform, highly relativistic outflow into the interstellar medium or a pre-explosion stellar wind \citep{RM05}. The non-uniformity of the ejecta being in the form of the velocity gradients results in the formation of mildly relativistic internal shocks inside the ejected gas while the interaction of the ejecta front with the ambient medium drives a highly relativistic external shock. Radiation produced at the internal shocks when they are above the photosphere is seen as the highly variable prompt emission in the gamma-ray and x-ray bands. Photons produced in the internal shock represent the delayed afterglow seen from the x-rays through the radio. Note, however, that there are alternative models of GRBs where the radiation emission can occur without formation of a shock, e.g., the Poynting flux model, which assumes that the energy is released and radiation is emitted at reconnection sites. Recent studies indicate that the Weibel instability, originally proposed in the framework of relativistic shocks, is also induced at reconnection sites \citep{Drake+08}. Thus, the Weibel-jitter theory discussed below may be relevant to both baryon-dominated and Pointing-flux-dominated ejecta.

\section{GRB magnetic fields}

GRB shocks are collisionless --- the collisional mean-free-path is orders of magnitude larger than the typical source size, therefore their microscopic structure must be determined by collective plasma phenomena that introduce effective collisions in the system via electromagnetic fields, which mediate interactions of particles in the absence of Coulomb collisions. On large scales --- much greater than the effective mean-free-path --- these shocks can be described in the hydrodynamic approximation. Two assumptions are conventionally made in the models of GRBs, namely that the magnetic fields are somehow generated and the electrons are heated up (or accelerated to form a non-thermal tail) at the shocks. 

Magnetic fields are generated at shocks by the Weibel instability \citep{ML99}. Numerical particle-in-cell (PIC) simulations \citep{Silva+03,Nish+03,Fred+04,Spit05,MSK06,CSA07}, that saturates its linear phase at a relatively low magnetic field, near the  equipartition with the electrons. At such low fields, protons keep streaming in current filaments, whereas the electrons, being much lighter than the protons, are quickly isotropized in the random fields and form a uniform background. Nonlinear evolution of the filaments leads to further amplification of the magnetic field up to $\epsilon_B\sim0.1$ on average (and $\epsilon_B\sim1$ locally in clumps) as one approaches the main shock compression. Interestingly, the fields at the shock front are strong enough that the radiative cooling of the electrons can be significant once they transit the front \citep{MS07,MS08}. However, weaker fields occupy a substantial region in the upstream region (foreshock) so that radiative cooling of the electrons may turn out to be substantial here. If so, one can expect the field strength to be anywhere between $\epsilon_B\sim0.001$ and $\sim 0.1$, depending on the actual profile of the field in the foreshock region. As of now, PIC simulations cannot reliably probe the structure of and radiative from the foreshock. However, a recent theoretical self-similar model of a foreshock field generation \citep{MZ08} predicts that strong foreshock fields are to be generated via the Weibel-like instability by shock-accelerated cosmic rays propagating away from it. The predicted spatial spectrum of the field at the shock front is 
$$
B_\lambda\propto\lambda^{-(p-2)/p},
$$
where $p$ is the power-law index of the cosmic ray population, typically $p\sim2.2$ for relativistic shocks. The energy density of the cosmic-ray generated field is of order of few percent of the cosmic ray kinetic energy density, which, in turn, can be as large as some tens percent of the shock kinetic energy density.

Since small-scale fields can decay relatively rapidly in the shock downstream, the Weibel-jitter model seems to be best applicable at the stage when the electron cooling is very fast, i.e., during the prompt and, possibly, the early afterglow phases. However, this limitation is relevant for the jitter emission mechanism, and not to the overall shock radiative efficiency. It is so because the cosmic-ray-generated fields are relatively large-scale and can sustain magnetic dissipation yet they are strong enough to ensure acceptable radiation efficiency even in the afterglow regime via standard synchrotron emission, at least, according to the simplified self-similar theoretical model. These multi-scale fields in around the shock front are needed for the efficient Fermi acceleration of radiating electrons and cosmic rays.

\section{ Electron cooling time. Relevance of PIC simulations} 

The radiative efficiency of a shock is determined by how fast the bulk electrons lose their energy via radiation. If the electron synchrotron cooling time is smaller than or comparable to the electron residence time in the high field region (near the shock front), this electron will radiate away energy comparable to its kinetic energy. The shock will be radiatively efficient in this regime regardless of the field dynamics in the far downstream. The dimensionless cooling time is defined as
$$
T_{\rm cool}=t_{\rm cool} \omega_{pp}
=\left(\frac{6\pi m_e c}{\sigma_T\gamma_e B^2}\right)
\left(\frac{4\pi e^2 n'}{\Gamma m_p}\right)^{1/2},
$$
where $t_{\rm cool}$ is the synchrotron cooling time, $n'$ is the particle density behind the shock measured in the downstream frame, $\Gamma$ is the shock Lorentz factor. The region of strong magnetic field at the shock front, where $\epsilon_B\sim 0.1 - 0.05$ is of the size $d_B \sim50 c/\omega_{pp}$ or so. Since the shock moves at $v=c/3$ in the downstream frame, the residence time in the region of high field is $t_{\rm res}\sim d_B/v\sim(150-300)\omega_{pp}^{-1}$. This estimate, does not account for the filling factor of magnetic inhomogeneities, which shortens the effective $t_{\rm res}$, and the electron trapping in high-field clumps, which is increasing $t_{\rm res}$.

If $t_{\rm cool}\la t_{\rm res}$, then the electrons lose their energy quickly near the shock jump, hence the radiative efficiency is high and such a shock can be seen as a GRB. We refer to a shock as the {\it ``radiative shock''} if $T_{\rm cool}\la300$ and as the {\it ``weakly radiative shock''} otherwise (its efficiency depends on the fields in the far downstream, which have not yet been adequately probed in PIC simulations). In an extreme case, $T_{\rm cool}<1$, called the {\it ``strong cooling''} regime, 
radiative cooling will be substantial even before the main shock compression, hence cooling may change the entire shock structure.

Using the standard shock model \citep{RM05}, the comoving density behind an internal shock at a radial distance $R$ from the central engine is
$$
n=4\Gamma_i L/(4\pi R^2 \Gamma^2 m_p c^3),
$$
where $L$ is the kinetic luminosity, $\Gamma_i$ is $\Gamma$ of an internal shock. 
The magnetic field and the electron bulk Lorentz factor are fractions $\epsilon_B$ and $\epsilon_e$ of the post-shock thermal energy density
$$
B'=\left(8\pi\Gamma_i m_pc^2n'\epsilon_B\right)^{1/2},
$$
$$
\gamma_e=(m_p/m_e)\Gamma_i\epsilon_e.
$$
Parameters $\epsilon_B$ and $\epsilon_e$ are $\sim10\%$ and $\sim50\%$ as follows from simulations. Finally, the dimensionless cooling time in baryon-dominated internal shocks becomes
$$
T_{\rm cool}^{(e^-p)}\simeq170 L_{52}^{-1/2}\Gamma_2\Gamma_i^{-3}
R_{12}\epsilon_B^{-1}\epsilon_e^{-1},
$$
Similarly, we evaluate the cooling time for the external (afterglow) shocks for the constant density ISM and the Wind ($n\propto R^{-2}$) outflow models \citep{GPS99,CL00}; the details are in
\citep{MS08}.

\begin{figure}
\includegraphics[height=0.32\textheight]{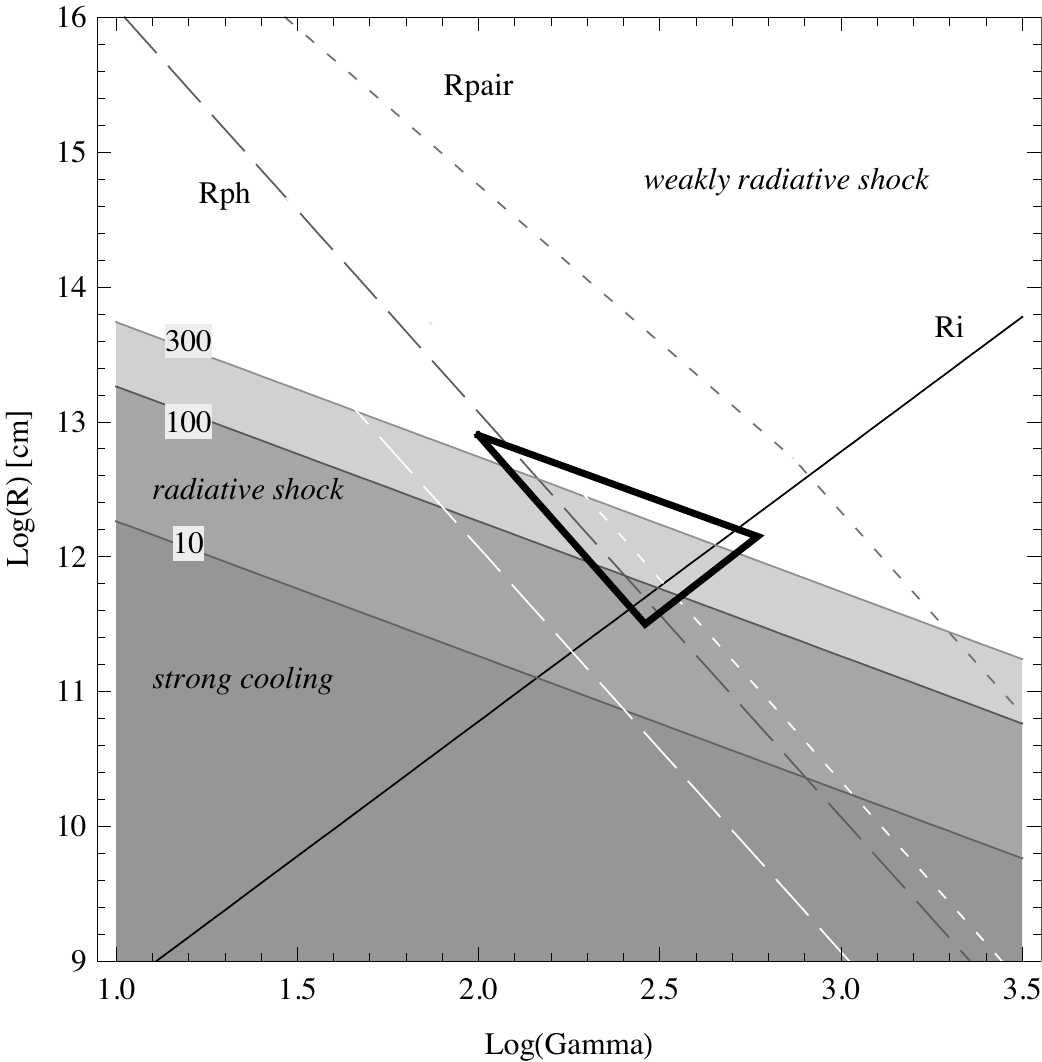}~~~%
\includegraphics[height=0.32\textheight]{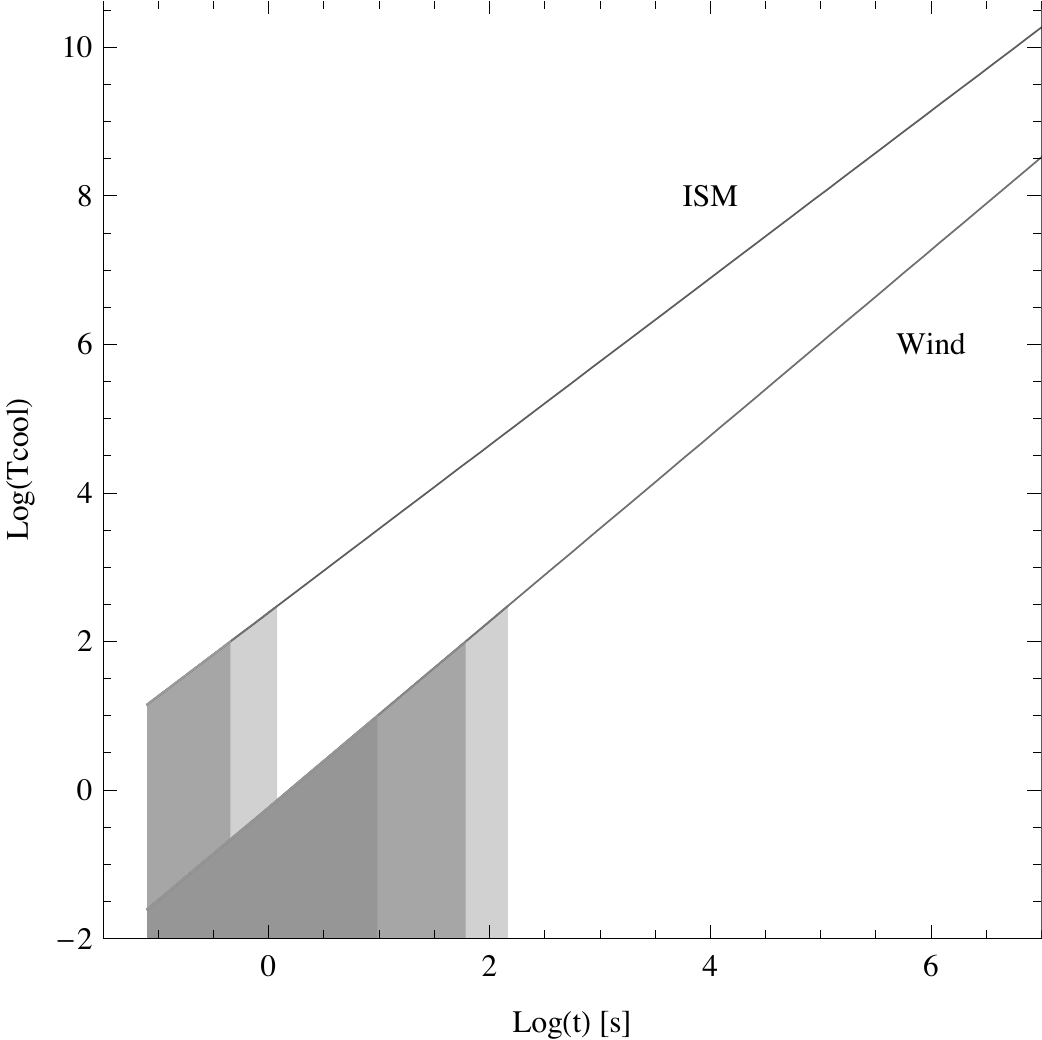}
\caption{ ({\it left}) --- Contours of $T_{\rm cool}$ vs. $\Gamma$ for the internal shocks for $T_{\rm cool}=1,\ 100,\ 300$. Dark filled regions correspond to $T_{\rm cool}<1$ (strong cooling), medium-dark and light regions correspond to $1<T_{\rm cool}<100$ and $100<T_{\rm cool}<300$, respectively (radiative shock), and the white region corresponds to $T_{\rm cool}>300$ (weakly radiative shock). Here $\Gamma_i=4$, $L_{52}=1,\ L_\gamma=0.1L,\ t_{v,-4}=1$. We also mark the radii beyond which the internal shocks can form ($R_i$), the medium is optically thin to Thompson scattering ($R_{\rm ph}$) and the optical depth due to $e^\pm$-pairs is below unity ($R_{\rm pair}$).
({\it right}) --- 
Cooling time in afterglows vs. time after the burst, for the ISM and Wind models. We use, $E=10^{54}$~erg, $A_*=10$ and $n_{\rm ISM}=100~{\rm cm}^{-3}$ and  a typical $z=2$.
\label{f1}}
\end{figure}

The results are shown in Fig. \ref{f1}. One can see that internal shocks ($\Gamma_i\sim4$) with strong foreshock emission, $T_{\rm cool}\la1$, can occur for $\Gamma\la60$ and at $R\la{\rm few \times10^{10}-10^{11}}$, well below the photosphere at such low $\Gamma$'s. Strongly radiative shocks
$T_{\rm cool}\sim100-300$ can occur above the baryonic photosphere in a relatively narrow, but very natural range of parameters,  $\Gamma\sim150-400$, $\Gamma_i\ga2.5$, $R_i\sim10^{12\pm0.5}$cm and the outflow kinetic luminosity $L\sim10^{52\pm2}$erg/s; hence they are likely observable. Since $T_{\rm cool}\propto\Gamma_i^{-3}$, the region of the parameters  widens greatly with increasing $\Gamma_i$.

For external shocks, we also see that, except for the very early times, the afterglow emission should be coming from far downstream, not from the main shock compression region. However, in the Wind model, the external shock can be radiative up to $\sim100$~s after the burst, whereas for the ISM model, the radiative shock regime can occur only at times earlier than a second after the explosion. Since the afterglow usually sets in at least several tens of seconds after the explosion (when enough external gas is swept by the shock), we conclude that very early afterglow emission can, in principle, come from radiative shocks propagating in relatively strong Wolf-Rayet  winds ($A_*\ga$~few).

It is remarkable that under these conditions, all the emission shall come from a thin shell of thickness $\sim t_{\rm cool}c\sim1$ meter (internal shocks) and $\sim100$ km (external shocks), that is, from the region of main shock compression. This region is already well resolved in 2D PIC simulations. Moreover, one can obtain the emitted radiation directly from PIC simulations.

\section{Radiation: spectral evolution and spectral correlations. Observational predictions}

The radiation calculation is tricky in the {\em Weibel model}, where the generated B-fields have the coherence length much smaller than the Larmor radius --- hence, {\em jitter} radiation is emitted \citep{M00}. It's been demonstrated \citep{M06} that the jitter radiation field is anisotropic with respect to the direction of the Weibel current filaments and that its spectral and polarization characteristics are determined by microphysical plasma parameters. In particular, the peak frequency differs from the synchrotron one: $\nu_{jitter}\simeq(\epsilon_B/10^{-3})^{1/2}\nu_{synch}$, see \citep{M+07}. Using the parameters of PIC simulations discussed above, we predict that radiatively efficient shocks shall have a spectral peak, $E_p$ in the MeV range, likely from tens to few hundred MeV. GLAST will be able to tell us whether the radiatively efficient shocks, which shall form near or just outside the photosphere, form in GRBs. 
However, if most of the emission is produces in the extended foreshock, where  $\epsilon_e$ and $\epsilon_B$ are lower, rather than at the main shock jump (as we assumed in the analysis above), then $E_p$ will be close to the conventional values of few tens to few hundred keV.

\begin{figure}[t!]
\includegraphics[height=0.24\textheight]{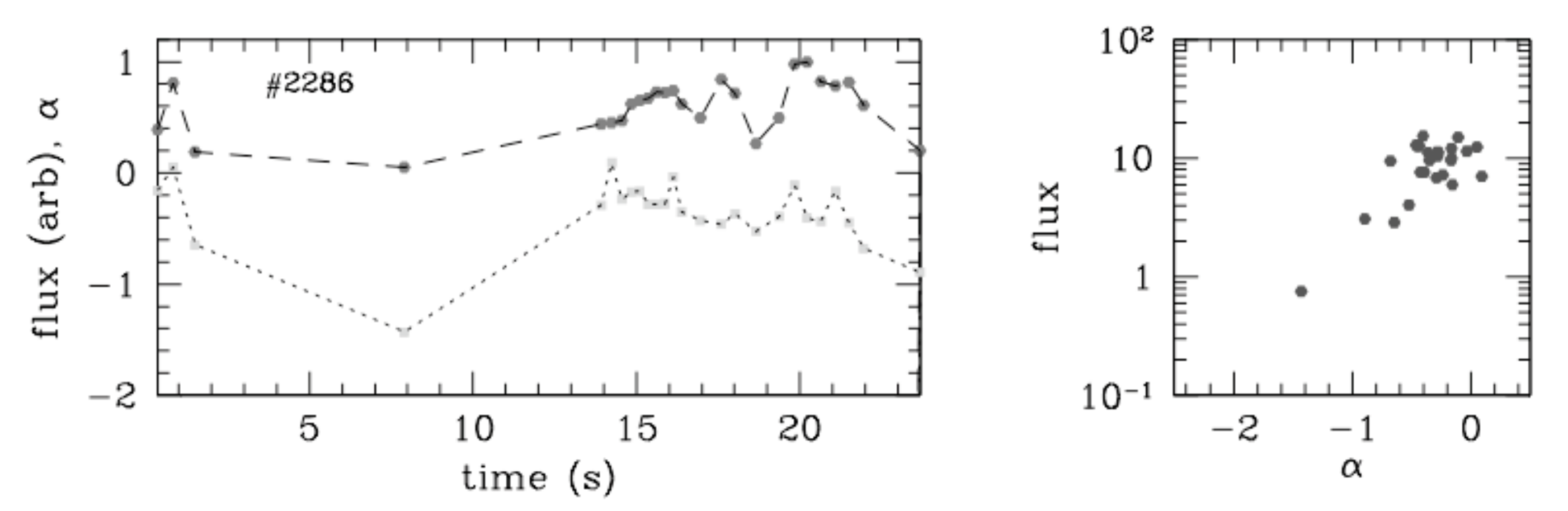}
\caption{ An example of the burst with tracking behavior (BATSE burst 2286). Left panel shows the normalized prompt light-curve (upper curve) and the time dependence of the low-energy photon spectral index $\alpha$ (lower curve). Right panel is the scatter plot of the flux and $\alpha$ highlighting the correlation of the spectral parameters. 
\label{f2a}}
\end{figure}

\begin{figure}[b!]
\includegraphics[height=0.3\textheight]{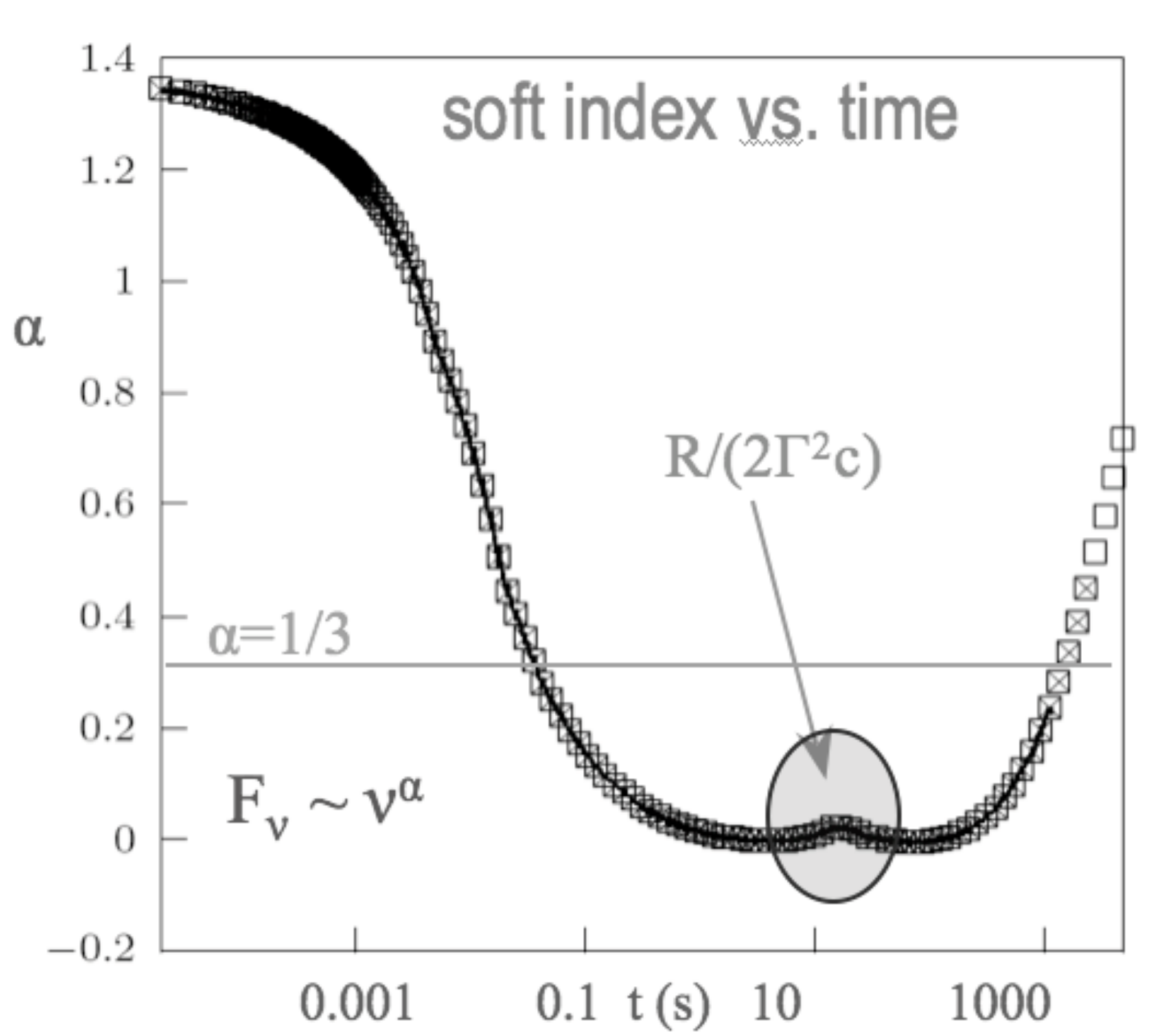}
\includegraphics[height=0.3\textheight]{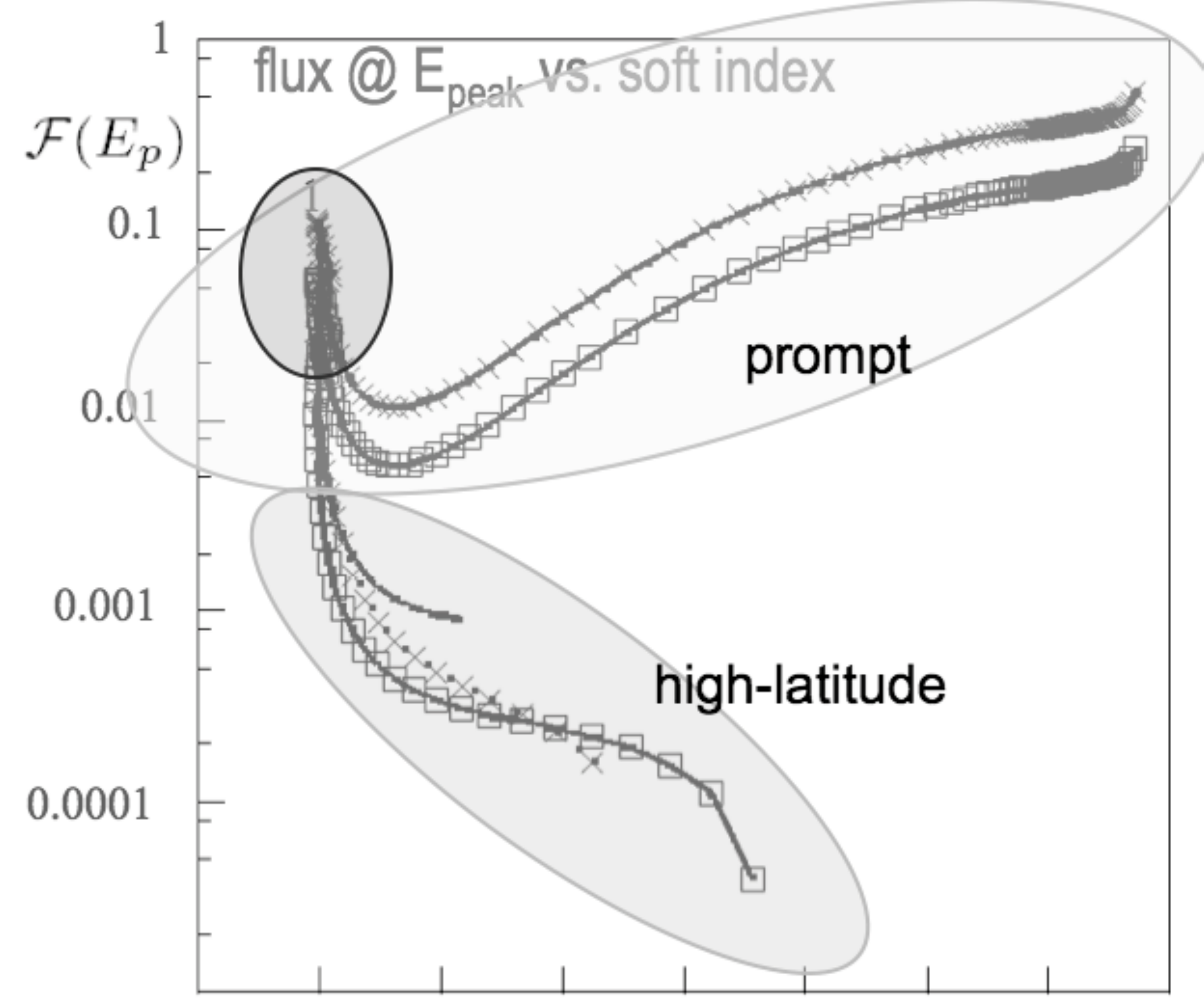}
\caption{ ({\it left}) --- Evolution of the soft spectral power-law index as a function of Log(time) for an individual emission episode (a  single pulse). Initially, the spectrum is extremely hard $\alpha\sim 1.4$ (harder than the synchrotron low-energy limit $\alpha=1/3$ shown with a horizontal line).  The spectral softening occurs on the $R/2c\Gamma^2$ time-scale and the photon spectral index evolves to the values a little below zero.
({\it right}) --- Correlation of the normalized flux at $E_p$ and $\alpha$ (the horizontal axis runs from $\alpha=-0.2$ to $\alpha=1.4$). The positive and negative correlations during the prompt and the high-latitude phases respectively are identified with large ovals. The positive correlation is observed in some GRBs \citep{Kaneko+06}, the negative one is the prediction of the model. The region marked in both panels with a small oval shows when a polarization signal, if any, may be expected. 
\label{f2}}
\end{figure}

Quite a large number of prompt GRBs exhibits tracking behavior, when the low-energy spectral index is correlated the observed gamma-ray flux --- the feature which is difficult to reconcile with the standard synchrotron shock model. Such a behavior is illustrated in Figure \ref{f2a}.
It's been shown \citep{M06} that the flux-$\alpha$ correlation ($\alpha$ being the soft spectral index) and the tracking pattern can be a combined effect of temporal variation of the shock viewing angle and relativistic aberration of an instantaneously illuminated, thin shell. Preliminary results are presented in Figure \ref{f2}. The model predicts that hard (e.g., synchrotron violating, with $\alpha\ge1/3$) spectra result from a shock patch close to the line of sight and that they are associated with the onset of a sub-pulse. [Note, here we use denote $\alpha$ as the $F_\nu$ soft index, not the photon index ($\alpha_{ph}=\alpha-1$) traditionally used in Band function fits.] 

Figure \ref{f2} (left panel) represents the evolution of the soft spectral power-law index as a function of time in logarithmic scale for an individual emission episode --- a single sub-pulse within a GRB prompt burst. At early times, the spectrum is extremely hard $\alpha\sim 1.4$ which is much harder than the synchrotron steepest low-energy spectrum ($\alpha=1/3$ shown as a horizontal line). 
The spectral softening occurs on the $R/2c\Gamma^2$ time-scale and the photon spectral index evolves to $\alpha\sim 0$ or even to a little below zero. Most of the emission of a GRB is, thus, softer than synchrotron. This naturally explains why the peak of the distribution of $\alpha$ is at $\alpha\sim 0$ as obtained in a comprehensive time-resolved spectral analysis of the BATSE GRB catalog data \citep{Kaneko+06}. The presence of a low-energy break in the jitter spectrum at oblique angles also explains the appearance of a soft X-ray component in some GRBs \citep{M06}. 

The spectral evolution of $\alpha$ is co-temporaneous with the variation of the observed photon (or energy) flux as being due to the variation of the Doppler boost in the observed frame. Thus, there shall be a correlation of the above parameters. Figure \ref{f2} (right panel) represents the dependence (correlation) of the flux at $E_p$ and $\alpha$. The prompt and the high-latitude phases are identified with large ovals. The positive correlation during the prompt phase seen in the figure is observed in a number of GRBs \citep{Kaneko+06}. So far this correlation has not explained with any other mechanism. At late times, the negative correlation (i.e., hardening of the low-energy part of the spectrum) is predicted by our model. Shall such a spectral trend in the high-latitude emission phase be observed, this will strongly support the applicability of the Weibel-jitter model. 

At last, the region marked in both panels with a small oval corresponds to the shock seen nearly edge-on. The magnetic fields, random yet lying in the plane of the shock, are also seen edge-on which may result in partially polarized radiation. Therefore, a polarization signal, if any, may be expected at this time, if the symmetry of the jet is not perfect, e.g., a jet axis is misaligned with the line of sight.

\begin{figure}[t!]
~~~~~~~~~~~\includegraphics[height=0.35\textheight]{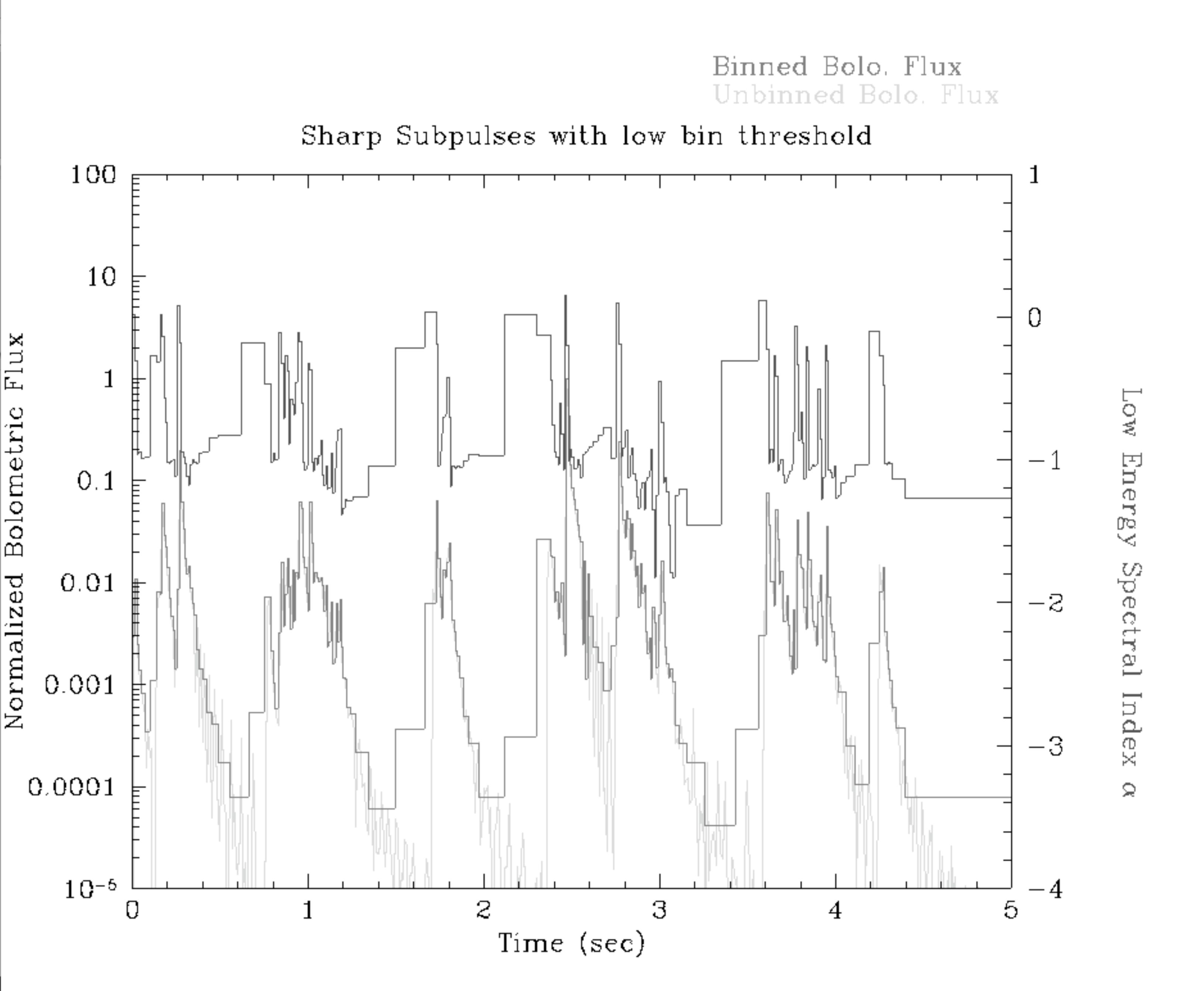}
\includegraphics[height=0.35\textheight]{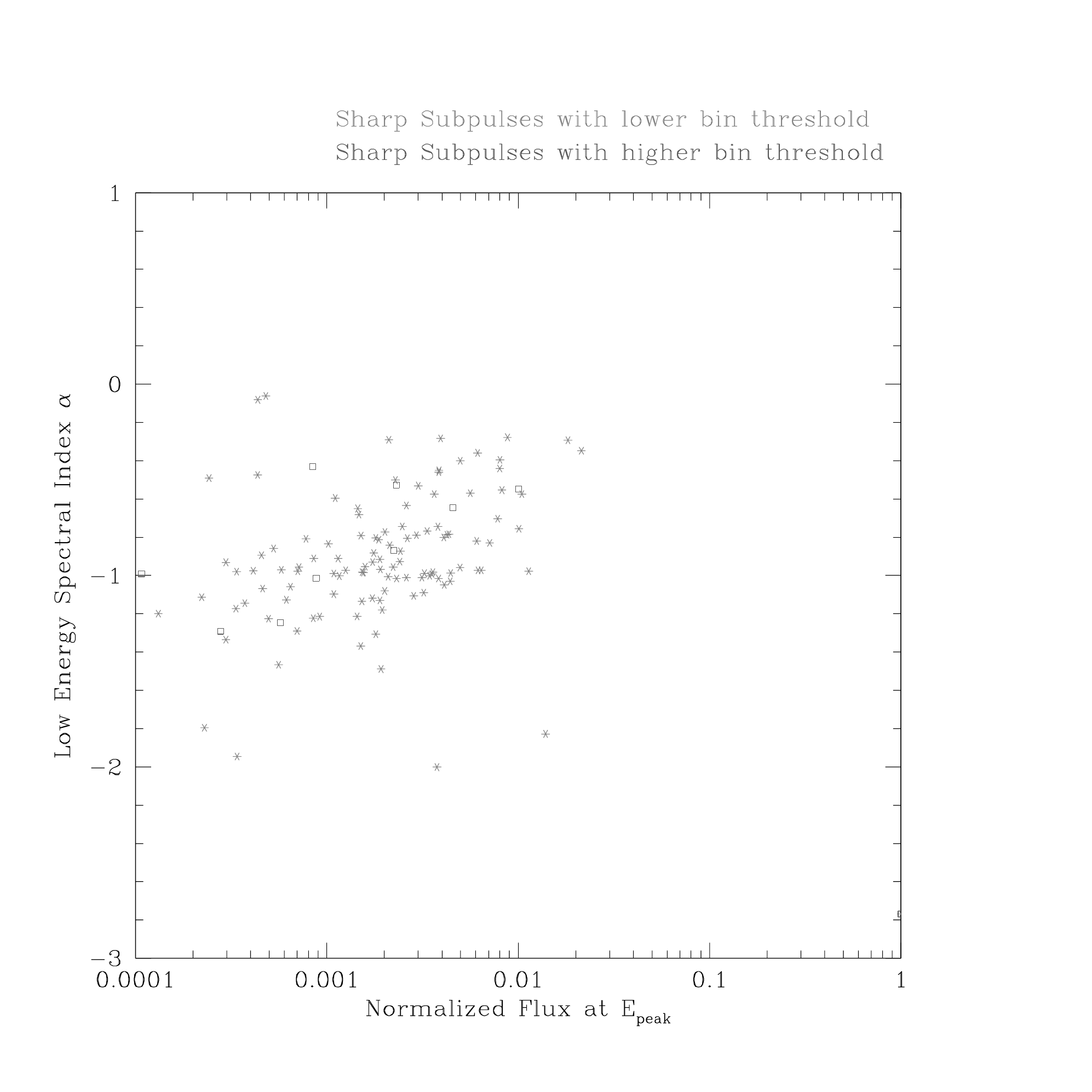}
\caption{ ({\it left}) --- Synthetic light-curve of the fluch at the peak energy (original and binned in constant fluence time bins -- lower curves) and the computed dependence of the soft photon spectral index $\alpha$.  
({\it right}) --- Correlation of the normalized flux at the peak energy and $\alpha$ for a high and low fluence bin threshold (which mimics bright and dim bursts).
\label{f2b}}
\end{figure}

At last, we made a synthetic GRB and analyzed it using Band function. The light-curves and the scatter plot of the spectral parameters are shown in Figure \ref{f2b}. The tracking behavior and the spectral correlation similar to those found in BATSE time-resolved bursts is clearly evident.

At the end, we mention that predictions of the Weibel-jitter model for afterglows have also been made \citep{M+07}: one expects a softer-than-synchrotron spectrum, with the photon index below the synchrotron break and above the self-absorption break being around $-1$. The positions of the breaks also differ from those in the synchrotron model.

\begin{theacknowledgments}
This work was supported by grants AST-0708213 (NSF), NNX08AL39G (NASA), DE-FG02-07ER54940 (DOE). The author thanks S. Pothapragada and S.Reynolds for their valuable contribution to the Radiation section.
\end{theacknowledgments}

\end{document}